\documentstyle[12pt]{article}

\newcommand{\bx}[1]{\mbox{\boldmath $#1$}}
\catcode`\@=11
\def\eqalign#1{\null\,\vcenter{\openup\jot\m@th
\ialign{\strut\hfil$\displaystyle{##}$&$\displaystyle{{}##}$\hfil
     \crcr#1\crcr}}\,}
\catcode`\@=12
\def\QBP{\,\overline{\! Q'}}

\begin{document}

\title{Tests of flavour independence
 in heavy quark potential models
}
\author{Jean-Marc Richard\\
{\sl Institut des Sciences Nucl\'eaires}\\
{\sl Universit\'e Joseph Fourier, CNRS-IN2P3}\\
{\sl 53, avenue des Martyrs, 38026 Grenoble Cedex, France}
}
\maketitle
\begin{abstract}
We review   some rigourous consequences of flavour independence
 on the spectrum and
properties of hadrons in potential models, with emphasis on hadrons
with two
heavy quarks, such as $(b\bar c)$ mesons and  $(QQq)$ baryons.
\end{abstract}
%

\section{Introduction}
Potential models are rather successful in
hadron spectroscopy.  Once the  spectrum of known states  is
reproduced with a
few parameters, some predictions can be made.
For instance, the location of the P-states of  charmonium or
bottomonium
 has been guessed from the lowest S-states and their radial
excitations; the
mass of the $\Lambda_b(bud)$
 baryon was predicted near $5.6\,$GeV in any reasonable potential
model, while
a
much larger mass was sometimes computed in other approaches;  a
recent
survey of realistic potentials gives  the mass of the
lowest $(b\bar c)$ state  with an uncertainty as small as $\pm
20\,$MeV; etc.

In the light-quark sector, the success of potential models is a
little
accidental, though one can argue that the Schr\"odinger equation
generates a
spectrum which is a very regular function of the constituent masses
and thus
mimicks the regularities of the genuine QCD spectrum in flavour
space.

For heavy  quarkonia $(Q\QBP)$ or heavy baryons $(QQ'q)$ with two
heavy quarks,
there is a deeper reason. The true picture results in a potential for
describing the relative motion of the two heavy quarks, once the
gluon and
light-quark degrees of freedom are integrated out. This is similar to
the
Born--Oppenheimer treatment of the inter-nuclear motion in molecular
physics.
The
potential between the heavy quarks has first been guessed in
phenomenological
works, and then derived from more fundamental studies.

A key property of QCD is flavour independence. The gluons are coupled
to the
colour of the quark  independently of its isospin, hypercharge, or
charm. This
means the potential is the same whatever quarks experience it, at
least before
any relativistic correction is included. This has motivated studies
on
how the 2-body or 3-body Schr\"odinger bound states evolve  when the
constituent masses vary in a given potential. Some results will be
summarized
below. Amazingly, these studies have found applications in atomic
physics,
where
we have a similar situation, namely the very same Coulomb potential
binding
masses as different as $e^+$, $\mu^+$ or $p$.
\section{Some early results on flavour independence} \label{se:early}
\subsection{Gap to the OZI-allowed threshold} \label{subse:OZI}

In the simplest model, mesons are given by the
Hamiltonian
\begin{equation}
H(\alpha)=\alpha{\bf p}^{\;2}+V(r),
\end{equation}
where $\alpha$ is half the inverse reduced mass, and $V$ is
universal, i.e.,
independent of $\alpha$. The flavoured mesons
$(c\bar q)$ and $(b\bar q)$ have nearly the same
$\alpha\simeq1/(2m_q)$, and thus the same
energy. On the other hand,
\begin{equation}
\alpha(b\bar b)<\alpha(b\bar c)<\alpha(c\bar c),
\end{equation}
and since every level of $H(\alpha)$ has an
energy which increases with $\alpha$, we obtain
the hierarchy \cite{QR1,GM1}
\begin{equation}
G(b\bar b)<G(b\bar c)<G(c\bar c),
\end{equation}
where
\begin{equation}
G(Q\QBP)=(Q\bar q)+(\QBP q)-(Q\QBP)
\end{equation}
is the gap between the lowest quarkonium state $(Q\QBP)$ and its
Zweig-allowed
threshold.  Experimentally, $G(c\bar c)\simeq0.7\,$GeV, and $G(b\bar
b)\simeq1.1\,$GeV. %
\subsection{Quark-mass differences}\label{subse:q-m-diff}
The above considerations deal with the ground state of various
quark--antiquark configurations. Explicit models have shown that the
excitation spectrum of both $(c\bar c)$ and $(b\bar b)$ can be
simultaneously reproduced  by  the same potential. This is a
convincing
illustration of flavour independence. In such studies, the quark
masses
are free parameters. There is some freedom in fixing their value, but
it
turns out that  quark-mass differences such as $(m_b-m_c)$ are rather
well constrained by the data. Shifting both $m_b$ and $m_c$ up or
down
mostly results in changing the size of the wave functions.
Leptonic  or radiative widths can  thus be used to choose an optimal
set
of values. Rigourous bounds on $(m_b-m_c)$ from the spectrum have
been
studied by Bertlmann and Martin \cite{BeMa}.
\subsection{{SU(3)$_{\rm F}$} breaking for baryon
masses} \label{subse:su3}
If one dares at applying  non-relativistic
 models to baryons consisting of light quarks, the
flavour-independent character of  the central potential provides
convexity relations such as
\begin{equation}
\eqalign{
(qqq)+(qss)&<2(qqs)\cr
(sss)+(sqq)&<2(qss)\cr}
\end{equation}
Hyperfine
corrections contain  $1/(m_im_j)$ factors which makes them larger for
$q$ than for $s$ quarks. This compensates the above inequalities in
specific combinations, and one eventually obtains a simple
understanding
of the famous Gell-Mann--Okubo formula and ``equal-spacing'' rule of
decuplet
 \cite{RTannals}
\begin{equation}
\label{GMO-spacing}
\eqalign{
2(N+\Xi)&\simeq3\Lambda+\Sigma\cr
\Omega-\Xi^\star&\simeq\Xi^\star-\Sigma^\star
\simeq\Sigma^\star-\Delta\cr
}
\end{equation}
\subsection{Mass of {$\Lambda_b(bdu)$}}
\label{subse:lambdab}
Consider the Hamiltonian
\begin{equation}
H={{\bf p}_1^{2}\over2m_1}+{{\bf p}_2^{2}\over2m_2}+{{\bf
p}_3^{2}\over2m_3}
+V({\bf r}_1,{\bf r}_2,{\bf r}_3),
\end{equation}
where $V$ is a given operator, not necessarily pairwise
(one could even accept
here some relativized form of the kinetic energy for the quarks 2 and
3).
The lowest binding energy is an increasing and concave function
of the inverse
mass $m_1^{-1}$. This leads to an upper bound on the binding of
$\Lambda_b(bdu)$, when extrapolated from $\Lambda(sud)$ and
$\Lambda_c(cud)$.
The inequality involves quark masses, which are not observable,
but cannot
be varied beyond a limited range. The study can be refined to
accommodate
spin--spin forces, and supplemented by a lower bound on $\Lambda_b$
in terms of  meson masses, if one
assumes  a certain relation between the quarkonium and the baryon
potentials,
on which more shortly.
The mass of $\Lambda_b$ is eventually  constrained
in a rather narrow window
near $5.6\,$GeV, and explicit estimates are indeed,
clustered around this value \cite{MaRi}.
\section{Applications to {$(b\bar c)$}}
\label{se:bcbar}
At first sight, one expects the lowest $(b\bar c)$ meson
approximately half between $J\!/\Psi$ and $\Upsilon$. In a
flavour-independent
potential, this is in fact a lower bound \cite{BeMa},
i.e., we have
\begin{equation}
\label{bc-cc-bb}
2(b\bar c)\ge (c\bar c)+ (b\bar b).
\end{equation}
If one knows the excitation spectrum of $(c\bar c)$ and $(b \bar b)$,
one can extract model-independent bounds on the average kinetic
energy in the ground state, which governs the evolution of  the
ground-state
energy when
the reduced mass varies. This leads to an upper bound on
the lowest  $(b \bar c)$  state \cite{Bagan},
\begin{equation}
\label{bc-upper2}
\eqalign{
( b\bar c)\le{(c\bar c)+( b\bar{b})\over2}&-{9\over8}\delta E(c\bar
c)\left[1-
\left({m_ b+m_c\over 2m_{
b}}\right)^{1/3}\right]\cr
 {}&+2\delta E( b\bar b)\left[
\left({m_ b+m_c\over 2m_{c}}\right)^{1/3}-1\right],\cr
}
\end{equation}
where $\delta E$ denotes the orbital excitation energy,
$E(\ell=1)-E(\ell=0)$.
In summary,
\begin{equation}
6.26\le(b\bar c)\le6.43\,{\rm GeV}/c^2,
\end{equation}
for the spin-averaged ground state,
 and, indeed,
all predictions of realistic potentials  cluster near 6.3
GeV/$c^2$ \cite{Eichten-Quigg-bc}, in between the lower and the upper
bounds
provided by flavour
independence. With hyperfine corrections, one obtains typically 6.26
GeV/$c^2$ for the pseudoscalar, and 6.33 GeV/$c^2$ for the vector
\cite{Eichten-Quigg-bc}.
\section{Baryons with two heavy quarks}
\label{se:QQq}
Regularity patterns similar to those of mesons are expected
 in the baryon sector
(the mathematics of the 3-body problem is of course more delicate
than
that of the 2-body one, and sometimes requires some mild conditions
on the shape of the confining potential, which are satisfied by all
current
models \cite{JMRrep}).  For instance, one expects an analogue of
(\ref{bc-cc-bb})
\begin{equation}
\label{cqq-ccq-qqq}
2(cqq)\ge (ccq)+(qqq)
\end{equation}
which leads to an upper bound $(ccq)\le 3.7\,$GeV for the centre of
gravity of
the
ground-state multiplet of $(ccq)$. A upper bound can also be derived
for
$(ccs)$.
On the other hand, the convexity relation
\begin{equation}
\label{bcq-ccq-bbq}
2(bcq)\ge (ccq)+(bbq),
\end{equation}
cannot be tested immediately, as well as the even more
exotic-looking \cite{Martin-bcs}
\begin{equation}
\label{bcq-bbb-ccc-qqq}
3(bcq)\ge (bbb)+ (ccc) + (qqq),
\end{equation}
and its analogue with $q\rightarrow s$.
Of more immediate use is the relation
\begin{equation}
\label{bcq-bqq-cqq-qqq}
(bcq)\ge (bqq)+(cqq)-(qqq),
\end{equation}
which leads to a rough lower bound $(bcq)\ge6.9\,$GeV$/c^2$, if one
inputs the
following rounded and spin-averaged values: $(bqq)=5.6$, $(cqq)=2.4$,
and
$(qqq)=1.1\,$GeV$/c^2$.

To derive these inequalities, one  uses the Schr\"odinger equation,
even for
the
light quarks. Very likely, the regularities exhibited by
flavour-independent
potentials also hold in more rigourous QCD calculations and
in the experimental spectrum. Any failure of the above inequalities
would
be very intriguing.

Sometimes, one can be more precise, and derive inequalities that
include
spin--spin corrections, for instance relations between $J^P=(1/2)^+$
baryons with different flavour content. See \cite{JMRrep} for
details.

Another mathematical game triggered by potential models
consists of writing inequalities among meson and baryon masses.
The basic relation is \cite{JMRrep}
\begin{equation}
\label{meson-baryon}
2(q_1q_2q_3)\ge (q_1\bar q_2) + (q_2\bar q_3) +(q_3 \bar q_1),
\end{equation}
obtained by assuming that the potential energy operators
fulfill the following inequality
\begin{equation}
\label{pot-mes-bar}
2V_{qqq}({\bf r}_1,{\bf r}_2,{\bf r}_3)\ge \sum_{i<j}V_{q\bar
q}(|{\bf
r}_i-{\bf r}_j|),
\end{equation}
which holds (with equality) for a colour-octet exchange, in
particular
one-gluon exchange, and for the simple  model
\begin{equation}
\label{string}
V_{q\bar q}(r)=\lambda r,\qquad V_{qqq}=\lambda \min_J(d_1+d_2+d_3)
\end{equation}
where $d_i$ is the distance from the $i$-th quark to a junction $J$
whose
location is adjusted to minimize $V_{qqq}$ \cite{Yshape}.
We already mentioned possible applications to $\Lambda$.
In the double-charm sector, we obtain
\cite{FR1} $(ccq)\ge 3.45\,$ GeV$/c^2$ for the $(1/2)^+$ state.
This is rather crude, not surprisingly. Years ago, Hall and Post
\cite{Post} pointed out in a different context
that the pairs are not at rest in a 3-body bound state, and that
their
collective
kinetic energy is neglected in inequalities of type
(\ref{meson-baryon}).

Computing the $(QQq)$ energies in a given potential model does not
raise
any particular difficulty. The 3-body problem is routinely solved by
means of
the
 Faddeev equations or variational methods.
On the other hand,  successful approximations  often shed some
light on the dynamics. In particular, the
Born--Oppenheimer method works very well
for large  ratios $(M/m)$ of the quark masses. At
fixed $QQ$ separation $R$, one solves the 2-centre
problem for the light quark $q$. The energy of $q$ is added to the
direct $QQ$
interaction to generate the effective potential $V_{QQ}(R)$ in which
the
heavy quarks evolve. One then computes the $QQ$
 energy and wave function. Note that one can remove
the centre-of-mass motion exactly, and also estimate the hyperfine
corrections.

The physics behind the Born--Oppenheimer approximation is rather
simple.
 As the heavy quarks move slowly, the light degrees of freedom
readjust
themselves to their lowest configuration (or stay in the same $n$-th
excitation,
 more generally).
At this point, there is no basic difference with quarkonium.
The $Q\,\overline{\!Q}$
potential does not represent an elementary process. It can be viewed
as
the effective
interaction generated by the gluon field being in its ground-state,
for a given
$Q\,\overline{\!Q}$ separation.

The results shown in Table \ref{Table1} come from the simple
potential
\begin{equation}
\label{eq:potential}
V={1\over2}\sum_{i<j}\left[A+Br_{ij}^\beta+{C\over
m_im_j}\bx{\sigma}_i\cdot
\bx{\sigma}_j\delta^{(3)}({\bf r}_{ij})\right],
 \end{equation}
with parameters $\beta=0.1$, $A=-8.337$, $B=6.9923$, $C=2.572$,
in units of appropriate powers of GeV.
The quark masses are $m_q=0.300$, $m_s=0.600$, $m_c=1.905$ and
$m_b=5.290\,$GeV.
The $1/2$ factor is a pure convention, although reminiscent from the
discussion
of inequalities (\ref{meson-baryon}) and (\ref{pot-mes-bar}).
The smooth central term
can  be seen as a handy interpolation between the short-range
Coulomb regime modified by asymptotic-freedom corrections
and an elusive linear regime screened by pair-creation effects.
The spin-spin term is treated at first order to estimate $M_0$.
This model fits all known ground-state baryons with at most one heavy
quark.
\begin{table}[h]
\caption{\label{Table1} Masses, in GeV,  of
{$(QQq)$} baryons in a simple potential
model. We show the spin-averaged mass  {$\,\overline{\!M}$},
and the mass  {$M_0$} of the lowest state with
 {$J^P=(1/2)^+$}.}
\begin{center}
\begin{tabular}{ccccccc}
State&$ccq$&$ccs$&$bcq$&$bcs$&$bbq$&$bbs$\\
$\,\overline{\!M}$&3.70&3.80&6.99&7.07&10.24&10.30\\
$M_0$&3.63&3.72&6.93&7.00&10.21&10.27\\
\end{tabular}
\end{center}
\end{table}

A more conventional Coulomb-plus-linear potential was used
in Ref.\ \cite{FR1}, with similar results. One remains, however,  far
from
the large number of models available for $(b\bar c)$
\cite{Eichten-Quigg-bc},
and the non-relativistic treatment of the light quark might induce
systematic
errors. The uncertainty is then conservatively estimated to be
$\pm50\,$MeV,
as compared to $\pm20\,$MeV for $(b\bar c)$.
Note also that the $b$-quark mass $m_b$ is tuned
to reproduce the experimental mass of $\Lambda_b$ at 5.62 GeV$/c^2$,
and this latter value is not firmly established.

The Born--Oppenheimer framework leaves room for improvements.
 A relativistic treatment of the light quark was attempted in
\cite{FR1},
using the bag model. For any given $QQ$ separation, a bag is
constructed in
which the light quark moves. The shape of the bag is adjusted to
minimize the energy. In practice, a spherical approximation is used,
so that the radius is the only varying quantity. The energy of the
bag and
light quark  is interpreted as the effective $QQ$ potential.
Unlike the rigid MIT
cavity, we have a self-adjusting bag, which follows the $QQ$ motion.
Again, this is very similar to the
bag model picture of charmonium \cite{Kuti}.

Unfortunately, there are variants in the bag model, with different
values
of the parameters, and with or without corrections for
the centre-of-mass motion. These variants lead to rather different
values
for the $(ccq)$ masses \cite{FR1}.
This contrasts with the clustered shoots of potentials models,
and deprives the bag model
of predictive power in this sector of hadron spectroscopy.

It is hoped that the $QQ$ potential will be calculated by
lattice or sum-rule methods.

The excitation spectrum of $(QQq)$ baryons has never been calculated
in great
detail,
at least to our knowledge. In Ref.\ \cite{FR1}, an estimate is
provided for the
spin
excitation (ground state with $J^P=(3/2)^+$), the lowest
negative-parity level,
and
the  radial  excitation of the ground state.

The spin excitation is typically 100 MeV above the ground state, and
thus
should
decay radiatively, with an $M1$ transition. The orbital and radial
excitations
of
$ccq)$ are unstable, since they can emit a pion. The radial
excitation of
$(ccs)$ can
decay into $(ccq)+K$, but the orbital excitation cannot, and thus
should be
rather
narrow, since restricted to $(ccs)+\gamma$, or to the
isospin-violating
$(ccs)+\pi^0$.
\section{Exotic hadrons ?} \label{se:exotics}
There are several types of multiquarks in the literature.
Jaffe's $H$ dibaryon with strangeness $S=-2$, $(ssuudd)$,
  or the ``Pentaquark'', ($(\bar csuud$, for instance) are
tentatively bound by chromomagnetic forces, while the ``Tetraquark''
 uses  a combination
 of flavour-independent chromoelectric forces, and Yukawa-type of
long range forces.

This latter contribution was pointed out by
 T{\"o}rnqvist \cite{Tornqvist} and Manohar and
Wise\cite{ManoharQQqq},
 who studied pion-exchange between heavy mesons, and stressed that,
among
others, some
$DD^\star$ and $BB^\star$ configurations experience attractive
long-range forces. By
itself, this Yukawa potential seems unlikely to bind $DD^\star$, but
might succeed
for the heavier $BB^\star$ system.

Years ago, Ader et al.\ \cite{AdRiTa} showed that $(QQ\bar q\bar q)$
should
become
stable for very large quark-mass ratio $(M/m)$, a consequence of the
flavour independence of chromoelectric forces. The conclusion was
confirmed in
subsequent studies \cite{Tetraq}.

In the limit of large $(M/m)$,  $(QQ\bar q\bar q)$ bound states
exhibit a
simple
structure. There is a localized $QQ$ diquark with colour $\bar 3$,
and this
diquark
forms a colour singlet together with the two $\bar q$, as in every
flavoured
antibaryon. In other words, this multiquark uses well-experimented
colour
coupling,
unlike speculative mock-baryonia or other states proposed in ``colour
chemistry''
\cite{Chan}, which contain  clusters with colour 6 or 8.

The stability of $(QQ\bar q\bar q)$ in flavour-independent potentials
is
analogous to
that of the hydrogen molecule \cite{Hydrogen2}.
If one measures the binding in units of the threshold energy, i.e.,
the energy
of two
atoms, one notices that the positronium molecule $(e^+e^+e^-e^-)$
with equal
masses
is bound by only 3\%, while  the very asymmetric hydrogen reaches
17\%. This
can be understood by writing
the molecular Hamiltonian as
\begin{equation}
\label{eq:hydrogen}
\eqalign{
H=&H_{\rm S}+H_{\rm A}\cr
=&\left({1\over4M}+{1\over4m}\right)\left({\bf p}_1^2+{\bf p}_2^2
+{\bf p}_3^2+{\bf p}_4^2\right)+V\cr
+&\left({1\over4M}-{1\over4m}\right)\left({\bf p}_1^2+{\bf p}_2^2
-{\bf
p}_3^2-{\bf
p}_4^2\right) }
\end{equation}
The Hamiltonian $H_{\rm S}$, which is symmetric under charge
conjugation, has
the
same threshold as $H$, since only the inverse reduced mass
$(M^{-1}+m^{-1})$
enters the
energy of the $(M^+m^-)$ atoms. Since $H_{\rm S}$ is nothing but a
rescaled
version
of the Hamiltonian of the positronium molecule, it gives 3\% binding
below the
threshold. Then the antisymmetric part $H_{\rm A}$ lowers the
ground-state
energy of $H$, a
simple consequence of the variational principle.

In simple quark models without spin forces, we have a similar
situation. The
equal
mass case is found unbound, and $(QQ\bar q\bar q)$ becomes stable,
and more and
more
stable, as $(M/m)$ increases.
One typically  needs $(bb\bar q\bar q)$, with $q=u$ or $d$, to
achieve binding
with the
nice diquark clustering we mentioned. However, if one combines this
quark attraction with the long-range Yukawa forces, one presumably
gets binding
for
$(cc\bar q\bar q)$ with $DD^\star$ quantum numbers.
A more detailed study is presently under way \cite{Tornqvist2}.

The experimental signature of tetraquark heavily depends on its exact
mass.
Above $DD^\star$, we have a resonance, seen as a peak in the
$DD^\star$ mass
spectrum.
Below $DD^\star$, one should look at $DD\gamma$ decay of tetraquark.
If it lies
below
$DD$, then it is stable, and decays via weak interactions, with a
lifetime
comparable
to that of other charmed particles.

\section*{Acknowledgements}
I would like to thank S.\ Narison
for the very stimulating atmosphere of this Workshop, and A.J.\ Cole
for useful
remarks on the manuscript.
%
%

\end{document}